\documentclass[preprint]{elsarticle}
\usepackage[utf8]{inputenc}
\usepackage[english]{babel}
\usepackage{amsmath}
\usepackage{multirow}
\usepackage{textcomp}
\usepackage{gensymb}

\usepackage{lineno}

\usepackage{setspace}
\begin{document}
\begin{frontmatter}

\title{Optimal cement paste yield stress for the production of stable cement foams}

\author[navier]{Blandine Feneuil}

\author[navier]{Nicolas Roussel}

\author[navier]{Olivier Pitois\corref{cor1}}
\ead{olivier.pitois@ifsttar.fr}

\cortext[cor1]{Corresponding author}

\address[navier]{Laboratoire Navier, UMR 8205, École des Ponts ParisTech, IFSTTAR, CNRS, UPE, Champs-sur-Marne, France}


\begin{abstract}

Production of morphology-controlled cement foams remains challenging, mainly due to bubble stability issues during cement setting. The use of cement paste with high yield stress is expected to promote stability by damping intrinsic bubble kinetics. Here we show however that for given W/C ratio, fresh foam stability can be achieved instead by decreasing the yield stress of the cement paste. Indeed, in this low apparent yield stress regime, van der Waals attraction between cement grains is reduced and grains are allowed to be efficiently packed by bubbles, providing enhanced mechanical properties. This result is obtained for two distinct additives used at controlled concentrations and without resorting to set accelerators, which highlights the general significance of the underlying stability mechanism. It offers a promising solution to produce stable cement foams at high air content.

\end{abstract}


\end{frontmatter}

\section*{Notations}
\begin{small}
\begin{tabular}{p{1.2cm} p{9cm}}
$R$ & Initial bubble radius in a foam\\
$\gamma$ & Air-liquid surface tension\\
$\rho_{liq}$ & Liquid density (1.0 g/cm$^3$)\\
$\rho_{c}$ & Cement density (3.2 g/cm$^3$)\\
$\Phi$ & Air volume content\\
$\Delta P$ & Pressure applied on cement paste during water extraction measurements\\
$\tau_{y,0}$ & Yield stress of reference cement paste, i.e. the unfoamed cement paste with the same water and additives content as the cement foam. This value is measured by spread test.\\
$\rho$ & Cement paste density\\ 
$W/C_i$ & Water to cement ratio in cement paste before mixing with precursor aqueous foam. \\
$W/C_f$ & Water to cement ratio in reference cement paste\\
$\tau_{y,0}^*$ & Critical value of reference cement paste yield stress above which cement foams are unstable\\
$\tau_{y,0}^{**}$ & Critical value of reference cement paste yield stress below which cement foams are unstable\\
$\tau_{y,foam}$ & Cement foam yield stress measured by start-of-flow experiment\\
$\tau_{y,aq}(\Phi)$ & Yield stress of aqueous foam calculated at air volume content $\Phi$. \\
$\tau_{y,int}$ & Yield stress of interstitial cement paste calculated from the cement foam yield stress $\tau_{y,foam}$.\\
$G_{foam}$ & Measured cement foam elastic modulus\\
$G_{aq}(\Phi)$ & Elastic modulus of aqueous foam calculated at air volume content $\Phi$. \\
$G_{int}$ & Elastic modulus of interstitial cement paste calculated from the cement foam elastic modulus $G'_{foam}$.\\
$\Phi_c$ & Maximal volume fraction of spheres, we take $\Phi_c=0.64$\\
\end{tabular}
\end{small}

\section{Introduction}

Cement foams are highly porous materials, which offer interesting thermal insulation pro\-per\-ties. When cement foam density decreases, thermal resistance is improved but mechanical strength decreases~\cite{2016_Samson}. The control of the pore morphology is a crucial issue to optimize all macroscopic properties at a given density. For example, pore size and pore connections are control parameters for acoustic absorption~\cite{2012_Hoang} and flow permea\-bi\-lity \cite{2018_Langlois}.

Several foam production methods have been reported in literature, including che\-mi\-cal foaming, air entrainment and mixing with aqueous precursor foam. In the precursor foam method, cement slurry and aqueous foam are prepared separately before being mixed together. The resulting cement foam morphology depends on (1) the precursor foam morphology, (2) the capacity of the mixing process to preserve the precursor foam bubble sizes and (3) the bubble size evolution in the sample at rest until cement setting. 
The third point is challenging. As long as the interstitial material is not solid, foam destabilization occurs through three mechanisms~\cite{2013_Cantat}: drainage is caused by density difference between air bubbles and cement paste, ripening is a gas transfer from smaller bubbles to bigger bubbles, and coalescence refers to thin film breakage between two neighbor bubbles. 
Both drainage and ripening are affected by the consistency of the suspending fluid, so a promising method to stop or slow them down is to increase the yield stress of the interstitial material~\cite{2010_Guignot,2014_Lesov,2014_Lesov_2}.

In the case of cement foams, yield stress of the interstitial paste results from attractive interactions between cement grains. It depends on both the intensity of these interparticle forces and particle volume content~\cite{2006_Flatt}. The latter is related to the water-to-cement ratio whereas interparticle forces can be tuned by additives \cite{2016_Gelardi_11}. For instance, superplasticizers adsorb on cement grains, which causes steric repulsion between cement grains and, at the macroscopic scale, the decrease of the yield stress~\cite{2012_Flatt}. 
In a cement foam made from precursor aqueous foam, surfactants are needed to ensure the stability of the films separating the bubbles. Some of them, mainly anionic surfactants, have been shown to have a strong affinity towards cement grains and change the yield stress of the constitutive paste~\cite{2017_Feneuil}. When they are added in small amount in cement paste, adsorbed molecules form an hydrophobic layer on cement grains and promote hydrophobic attraction, which leads to an increase of the macroscopic yield stress. On the other hand, when large amount of anionic surfactant is added to cement paste, surfactant molecules agglomerate into micelles or double layers on cement grains surface and leads to steric repulsion and a strong decrease of the yield stress \cite{2017_Feneuil}.

In this paper, we investigate the effect of cement paste yield stress on the stability of cement foams. The yield stress is controlled by using additives, either superplasticizer or appropriate amount of anionic surfactant. Additionally, in order to indirectly assess the rheological behavior of the cement paste as confined between the bubbles, we perform rheological measurements on the cement foams.

\section{Materials and methods}

\subsection{Materials}

Cement is a CEM I from Lafarge, from Lagerdorf factory. Specific surface provided by the manufacturer is 0.433~m$^2$/g. Chemical composition is given in table \ref{table_chimie_ciment_c4}.

\begin{table}[!ht]
\begin{center}
\begin{tabular}{|c c c c c c c c c|}
\hline
C$_3$S & C$_2$S & C$_3$A & C$_4$AF & CaO/SiO$_2$ & MgO & Na$_2$O +0.658 K$_2$O & SO$_3$ & Gypsum  \\ \hline  
60\% & 13\% &  2\% & 13\% & 3 & 0.8\% & 0.5 \%  & 2.5\% & 4\% \\ \hline 
\end{tabular}
\caption{Chemical composition of CEM I cement from Lafarge, Lagerdorf}
\label{table_chimie_ciment_c4}
\end{center}
\end{table}

\bigbreak

Two different surfactants are used.  Steol$^\text{\textregistered}$~270 CIT is an anionic surfactant provided by Stepan. Its molar mass indicated by the manufacturer is 382~g/mol and the active content is 68-72\%. Steol Critical Micelle Concentration (CMC) is 1.5~g/L in water and 0.3~g/L in cement pore solution~\cite{2017_Feneuil}. Triton\texttrademark ~X-100 (laboratory grade) is a non-ionic surfactant provided by Sigma-Aldrich; molar mass is 625~g/mol and CMC is 0.2~g/L in water and 0.15~g/L in cement pore solution~\cite{2017_Feneuil}.

\begin{figure}[!ht]
\begin{center}
\includegraphics[width=\textwidth]{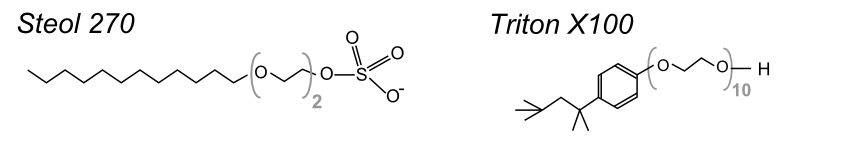}
\caption{Chemical formula of the surfactants used in this study: Steol is anionic and Triton is non-ionic}
\label{formule_Steol}
\end{center}
\end{figure}

When added into cement paste, Steol~270 has a strong affinity with cement grains surface~\cite{2017_Feneuil} and can either increase the cement paste yield stress at low concentration or act as a deflocculant at high concentration. On the contrary, non-ionic Triton has very low affinity with cement grains \cite{2001_Zhang,2017_Feneuil} and does not change the yield stress of the cement paste.

SIKA Tempo 12 superplasticizer has been used to modify cement paste yield stress in cement foam samples containing non-ionic surfactant. It has been checked that Tempo 12 does not alter the stability of aqueous foam made with Triton: we have compared the foam volume obtained by shaking tubes containing a Triton solution with and without superplasticizer. Results are not shown here, but details about the experimental method can be found in~\cite{2017_Feneuil}.

\subsection{Methods}

\subsubsection{Protocol}
\label{section_protocol_c4}

In order to remove any influence of cement paste age on results, all cement foams are prepared following the same time schedule, from initial water and cement mixing to sample casting or rheometry measurement. 

\begin{figure}[!ht]
\begin{center}
\includegraphics[width=\textwidth]{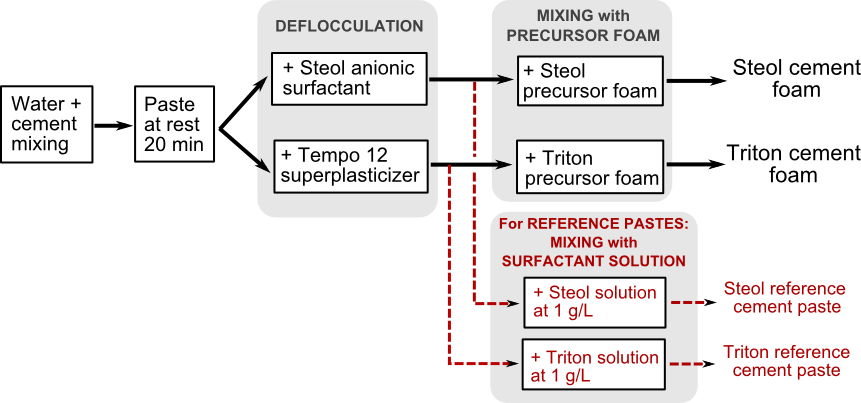}
\caption{Preparation protocol of cement foams and reference cement pastes made with Steol anionic surfactant and Triton non-ionic surfactant.}
\label{schema_protocol_c4}
\end{center}
\end{figure}

Water is mixed with cement paste at initial water-to-cement ratio $W/C_i$ = 0.35 or 0.32 and the resulting paste is left at rest for 20 minutes to allow for the formation of sulfo-aluminate phases. Then, a deflocculant is added. Deflocculant is either Steol surfactant added in large quantity or SIKA Tempo 12 superplasticizer. The resulting cement paste is then mixed with precursor aqueous foam. Final water-to-cement ratio after addition of additive and foam is noted $W/C_f$. This procedure is schematized in Fig.~\ref{schema_protocol_c4}.

In addition, reference cement pastes are prepared following the same protocol as cement foams, but only surfactant solution is added instead of aqueous foam (see Fig. \ref{schema_protocol_c4}). Reference cement pastes are then mixed slowly by hand to avoid any air entrainment.

\subsubsection{Precursor foam}

\begin{figure}[!ht]
\begin{center}
\includegraphics[width=10cm]{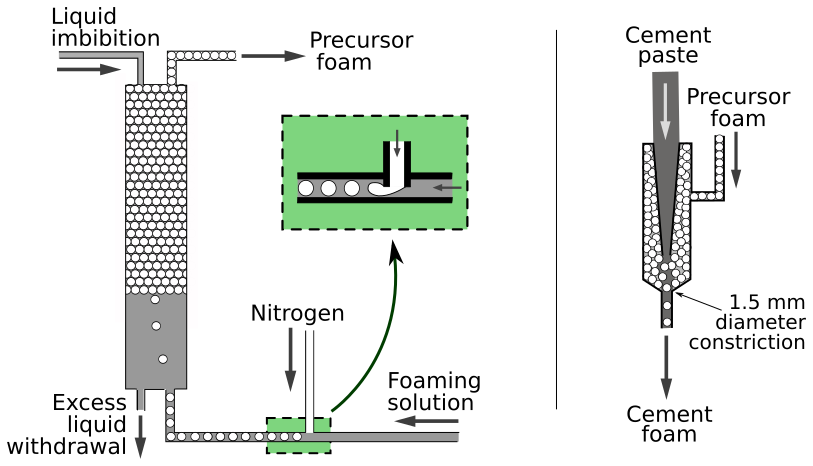}
\caption{Schema of precursor foam generation (left) and mixing device (right)}
\label{schema_FoamGeneration}
\end{center}
\end{figure}

The generator used to produce the precursor aqueous foam is schematized in Fig.~\ref{schema_FoamGeneration}. Nitrogen and surfactant solution (concentration 1~g/L for both surfactants) flow in a T-junction of characteristic length $l_T= 100~\mu$m. The capillary pressure $P_C$ depends on the gas-liquid surface tension $\gamma\sim 10~mN/m$ and is $P_C\sim\gamma/l_T \sim 10^2~Pa$. The hydrostatic pressure $P_H$ depends on the liquid density $\rho_{liq} \sim 1000~kg/m^3$: $P_H=\rho_{liq} g l_T \sim 1~Pa$. Therefore, as $P_C\gg P_H$, which means that capillary effects dominate over gravity effects, so liquid and gas pass alternately through the T-junction, which leads to the formation of bubbles. Bubble diameter depends on the flow rates of surfactant solution and nitrogen respectively.

The resulting bubbles are then collected in a vertical column. Imbibition flow at the foam top compensates liquid loss due to drainage and is used to tune the precursor foam liquid fraction. Note that, with such an approach, the vertical liquid fraction profile is uniform over the foam column. The precursor foam is mixed with cement paste about 40~minutes after the beginning of the generation, when ripening has not started to occur in the column.

Bubble radius for all precursor foams is $R=390\pm20$~µm and the liquid fraction is equal to 1.4$\pm$0.1\%. 

\subsubsection{Mixing}

To mix the precursor foam with the cement paste, we use a convergent mixing device schematized in Fig.~\ref{schema_FoamGeneration}. Cement paste flows in a conic tube, the length of which is 4~cm. Then, mixing of paste and precursor foam takes place thanks to a 1.5~mm diameter constriction. Input flow rates are chosen so that the final air content is $\Phi=83\%\pm 1\%$. Note that this mixing method involves the flow of the precursor cement paste in small tubes, which requires moderate cement paste yield stress (below about 40~Pa). 

For each cement foam sample, we fill both the rheometer cup for yield stress or elasticity measurement and a mold for stability assessment.

\subsubsection{Final stability}
\label{part_stability}

\begin{figure}[!ht]
\begin{center}
\includegraphics[width=9cm]{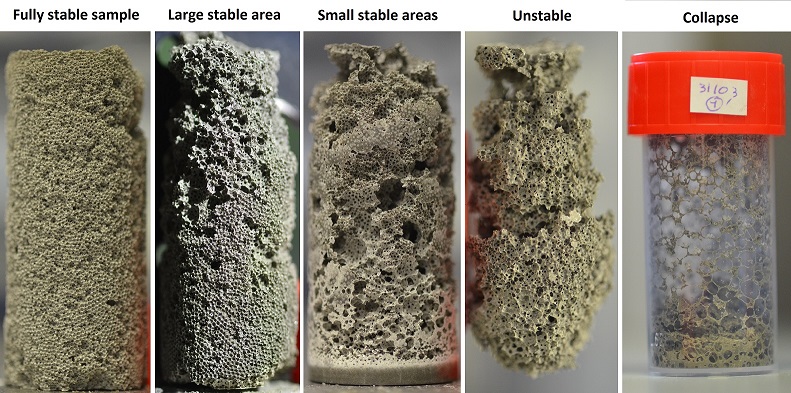}
\caption{Example of samples illustrating each stability class. Sample height is 6~cm.}
\label{schema_stability_scale}
\end{center}
\end{figure}

Samples are casted in 6~cm high 2.6~cm-diameter air-tight plastic cylinders. We checked that foams do no break when they are in contact with the mold walls. Samples are demolded 7 days after casting. Cement foam stability is assessed visually and samples are classified within the five categories illustrated in Fig.~\ref{schema_stability_scale}:

\begin{itemize}
\item 3: fully stable sample
\item 2: large stable area
\item 1: small stable area(s)
\item 0: no stable area
\item -1: sample collapse
\end{itemize}

It can be noted that the final stability is the result of many competing phenomena including the intrinsic stability of the foam and the time during which the foam is exposed to destabilisation. This time relates to the setting time of the system.

\subsubsection{Rheometry}

We use a stress controlled rheometer Malvern Kinexus Ultra+ with a Vane geometry in a striated (to avoid wall-slip) cylindrical cup of diameter 37 mm. The six-blade Vane tool is 5~cm high and 25~mm large. Each measurement sequence starts with stress relaxation during 30~s. Then, either yield stress is measured with a start-of-flow curve at a shear rate of 0.01~s$^{-1}$, or elastic modulus is monitored with 10$^{-5}$ oscillations at 1~Hz.

This strain value is well below both critical strains related to flocculation and formation of CSH bridges between cement grains~\cite{2012_Roussel}.  Therefore, we expect that elasticity measurement does not affect thixotropic behavior of the cement paste and is, as such, a non-destructive measurement.

\subsubsection{Water suction out of the cement pastes}

\begin{figure}[!ht]
\begin{center}
\includegraphics[width=6cm]{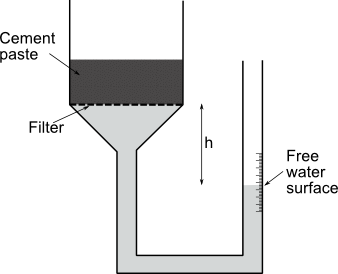}
\caption{Experimental device used to measure the water volume extracted from the reference cement pastes at $\Delta P \simeq 440$~Pa.}
\label{Schema_WaterExtraction}
\end{center}
\end{figure}

We measure the ability of the reference cement paste to release water by using the experimental device schematized in Fig.~\ref{Schema_WaterExtraction}. A 1.6-cm thick layer of reference cement paste containing Steol surfactant is placed on one side of a U-shaped tube filled with water. A filter (0.45~$\mu$m) separates the cement paste and the tube. It can be crossed by the water but not by the cement grains. The surface of water in the other branch of the tube is free to move. When the cement paste is raised above the free water surface at height $h$, a pressure difference $\Delta P =\rho_{liq} g h$ is created at the bottom of the cement paste. The volume of extracted water can then be deduced from the displacement of the free water surface in the tube. Note that $h$ decreases during the time of the experiment because water extraction makes the free water surface rise. For each experiment, the initial pressure is $\Delta P =$ 440~Pa and it decreases down to a value comprised between 400 and 375~Pa after 10~minutes. The value chosen for $\Delta P$ accounts for the hydrostatic pressure in the continuous phase in the 6-cm high foam samples described in paragraph \ref{part_stability} (where the pressure varies between 0 and 600~Pa). 

\subsection{Cement paste yield stress}
\label{Section_interstitial_Yieldstress}

To measure the reference cement paste yield stresses, noted $\tau_{y,0}$, a volume $\Omega$ (30 mL on average) of each paste is poured on a flat surface and the resulting average radius $R_{spread}$ is measured. $\tau_{y,0}$ is then obtained from the paste density $\rho$ and the poured volume $\Omega$ with the following formula~\cite{2005_Roussel}:

\begin{equation}
\tau_{y,0} = \dfrac{225 \rho g \Omega^2}{128 \pi^2 R_{spread}^5}
\label{equation_seuil_c4}
\end{equation}

Equation \ref{equation_seuil_c4} is valid at intermediate yield stress values. On the one hand, it must be high compared to capillary forces and, on the other hand, the sample thickness must be small compared to its radius. As already discussed in~\cite{2017_Feneuil}, capillary forces can be neglected when yield stress is above 1~Pa. We choose to set all the values measured below 1~Pa to $\tau_{y,0}=$ 1~Pa. Regarding the second condition, the maximal measured value is $\tau_{y,0} \simeq 100$~Pa: in this case, the ratio of the spread radius on the spread height is 3.

\begin{figure}[!ht]
\begin{center}
\includegraphics[width=6cm]{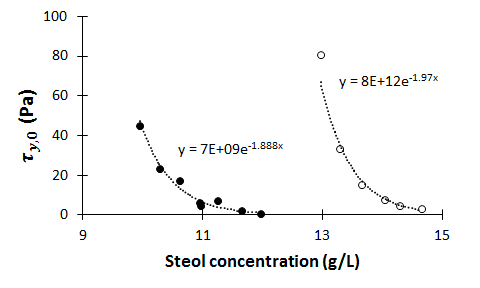}

\includegraphics[width=6cm]{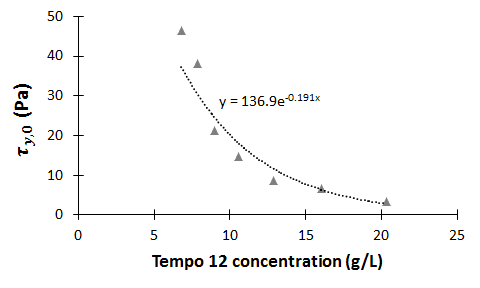}
\caption{Yield stresses of reference cement pastes prepared following the same protocol as cement foam. Top: pastes containing Steol anionic surfactant with W/C$_f$=0.38 (empty dots) and W/C$_f=$0.41 (full dots). Bottom: superplasticizer and Triton non-ionic surfactant with W/C$_f=$0.36. }
\label{graph_free_yieldstress}
\end{center}
\end{figure}

Measured values for the reference cement paste yield stresses $\tau_{y,0}$ are given in Fig. \ref{graph_free_yieldstress}. Two water-to-cement ratios W/C$_f=$ 0.38 and 0.41 have been studied in the case of Steol surfactant, whereas only W/C$_f=$0.36 has been considered in the case of Tempo 12 and Triton mixes. 

For the sake of simplicity, we choose here to interpolate our results with an exponential function as this mathematical form seems to fit our data over the range of our experiments. Therefore, in the following, we will use the following equations to estimate, when required, the reference cement paste yield stress $\tau_{y,0}$ from the Steol or Tempo 12 concentrations $C_{\text{Steol}}$ or $C_{\text{Tempo 12}}$ (with yield stresses in Pa and concentrations in g/L):
\begin{itemize}
\item $\tau_{y,0} = 6.87 \cdot 10^9 ~e^{-1.89 ~C_{\text{Steol}}}$ with Steol and W/C$_f$=0.41,
\item $\tau_{y,0} = 8.30 \cdot 10^{12}~ e^{-1.97 ~C_{\text{Steol}}}$ with Steol and W/C$_f$=0.38,
\item $\tau_{y,0} = 137 ~e^{-0.191 ~C_{\text{Tempo 12}}} $ with Tempo 12 and Triton.
\end{itemize}

\section{Results}

\subsection{Stability}

\begin{figure}[!ht]
\begin{center}
\begin{tabular}{c c c}
0 min & 10 min & 20 min\\
\includegraphics[width=0.29\textwidth]{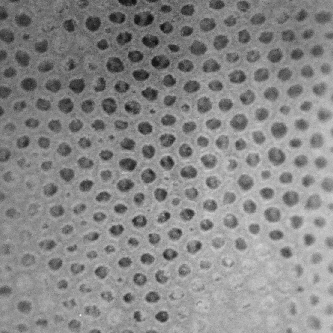} &
\includegraphics[width=0.29\textwidth]{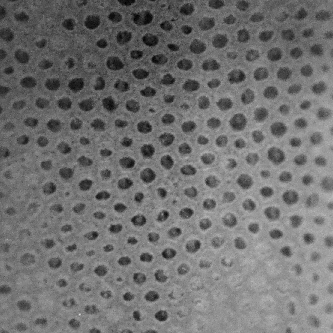} &
\includegraphics[width=0.29\textwidth]{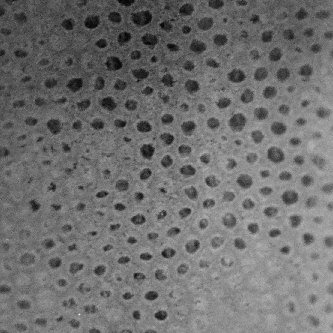} \\
30 min & 40 min & 50 min\\
\includegraphics[width=0.29\textwidth]{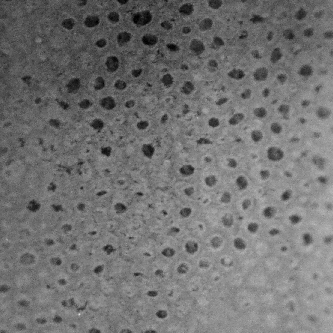} &
\includegraphics[width=0.29\textwidth]{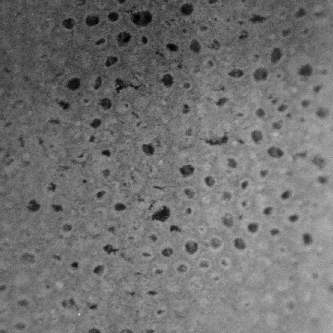} &
\includegraphics[width=0.29\textwidth]{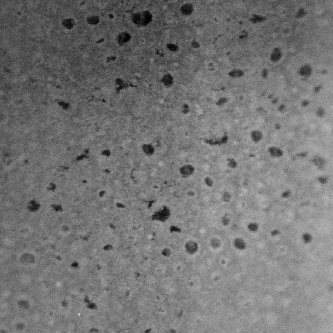} \\
\end{tabular}
\caption{Typical time evolution of an unstable cement foam ($W/C_f=0.41$, 10.2~g/L of Steol, $\tau_{y,0}=18~Pa$). Foam on the pictures is seen through the transparent mold ; note that the movement of the bubbles deposes a cement layer on the mold walls and that apparent morphology after foam destabilization may not be representative of the bulk morphology of the foam. Cement paste is in light grey and bubbles are black. Picture width is 1~cm.}
\label{pictures_destabilisation}
\end{center}
\end{figure}

Fig.~\ref{pictures_destabilisation} illustrates a typical time evolution of an unstable foam. Pictures are taken through the transparent mold, thus, note that we can see only the morphology of the side of the sample against the mold walls, where the movement of the bubbles deposes a cement layer. Therefore, the morphology of the foam, as seen at the wall after destabilization, may not be representative of the bulk morphology. However, we can neatly see that major change of the foam structure takes place before 30~min after sample preparation. This is true for all unstable samples, where destabilization always starts to occur before 30~min.

\begin{figure}[!ht]
\begin{center}
\includegraphics[width=6cm]{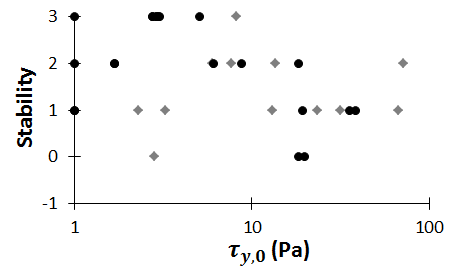}

\includegraphics[width=6cm]{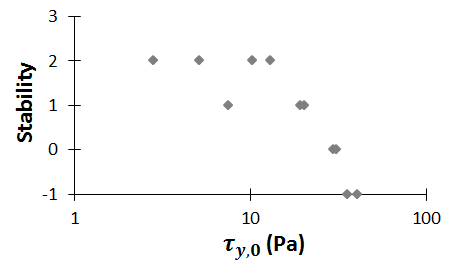}
\caption{Top: Stability of cement foams containing Steol anionic surfactant. Grey diamonds correspond to foams made with initial W/C$_f$=0.38 and black dots to foams with W/C$_f$=0.41. Bottom: Stability of cement foams containing superplasticizer and Triton. }
\label{graph_stability}
\end{center}
\end{figure}

The stability of cement foams samples containing either anionic or non-ionic surfactant is plotted in figure \ref{graph_stability} as a function of the reference cement paste yield stress $\tau_{y,0}$.

In all cases, best foam stability is obtained at moderately low reference cement paste yield stress. On the one hand, foams are unstable at high $\tau_{y,0}$, i.e. above a critical value $\tau_{y,0}^*\simeq 10$~Pa.  The increase of $\tau_{y,0}$ above $\tau_{y,0}^*$ leads consistently to unstable foams. This behavior is particularly noticeable in Tempo 12 - Triton foams, where foam collapse was observed for reference cement paste yield stress close to 25 Pa. On the other hand, very low reference cement paste yield stress values, below $\tau_{y,0}^{**} \simeq 2$~Pa, also lead to unstable foams. In this very low reference cement paste yield stress regime, reproductibility of the results is poor. Foams with the same formulation can sometimes have different stability behaviors.

These stability results are unexpected, because, as mentioned in the introduction, high interstitial yield stress is expected to stop drainage and ripening. These puzzling results can be due to several effects including:

\begin{itemize}
\item The cement paste inside the foam structure may have a different behavior than the unfoamed reference cement paste.
\item Time related effects, i.e. thixotropy and hydration kinetics, may play a significant role.
\end{itemize}

In the next paragraph, rheological investigation is used to elucidate these points. Foam yield stress measurement at early age is used to assess the behavior of the interstitial cement paste. Then, as a non-destructive measurement, elastic modulus is monitored to evaluate the time evolution of the paste rheological properties.
Besides, segregation of water and cement grains are measured by the water suction experiment.

\subsection{Rheological measurements}

\subsubsection{Foam yield stress}

\begin{figure}[!ht]
\begin{center}
\includegraphics[width=6cm]{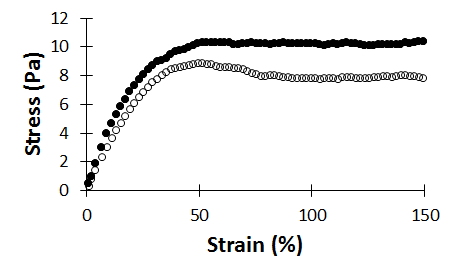}
\caption{Examples of the two types of start-of-flow curves. Both curves were obtained for samples with W/C$_f$=0.41, for two different Steol concentrations. Curve with overshoot (empty circles) is obtained with 11.4~g/L of Steol and $\tau_{y,0}=$3~Pa, and curve without overshoot (black dots), with 10.4~g/L of Steol and $\tau_{y,0}=$18~Pa.}
\label{graph_foam_start-of-flow}
\end{center}
\end{figure}

We notice that two types of start-of-flow curve shape were obtained for fresh cement foams. Examples for both curve shapes are shown in figure \ref{graph_foam_start-of-flow}. For some samples, the stress increases up to a plateau value. For other samples, plateau value is lower than the stress at the peak and the curve exhibits an overshoot. In both cases, yield stress $\tau_{y,foam}$ is obtained for shear strains between 40 and 75\%.

\begin{figure}[!ht]
\begin{center}
\includegraphics[width=7cm]{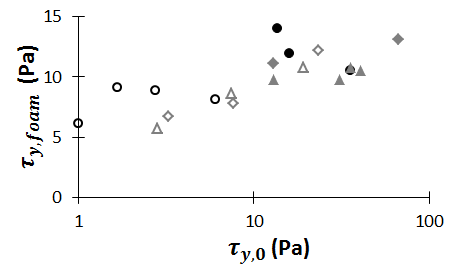}
\caption{Macroscopic yield stress of fresh cement foam. Black dots refer to cement pastes containing anionic surfactant at $W/C_f=0.41$ and grey diamonds at $W/C_f=0.38$. Grey triangles refer to samples with superplasticizer and non-ionic surfactant. Empty symbols refer to curves where an overshoot was observed.}
\label{graph_foam_yieldstress}
\end{center}
\end{figure}

Foam yield stress, $\tau_{y,foam}$, for all the samples is plotted as a function of the reference cement paste yield stress $\tau_{y,0}$ in Fig. \ref{graph_foam_yieldstress}. The presence of an overshoot is indicated by empty dots and diamonds.

For the three sets of results, $\tau_{y,foam}$ increases with $\tau_{y,0}$ but the foam yield stress variation is limited: while $\tau_{y,0}$ ranges from about 1 to 100~Pa, the foam yield stress varies from 6 Pa to 14 Pa. Besides, overshoot occurs when $\tau_{y,0}\leq\tau_{y,0}^*$.

\subsubsection{Elasticity}

\begin{figure}[!ht]
\begin{center}
\includegraphics[width=6cm]{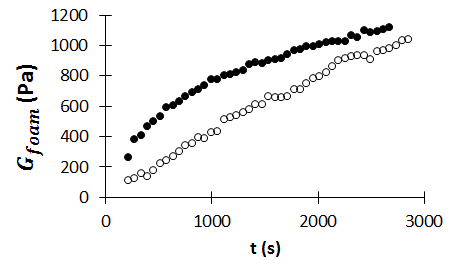}
\caption{Examples of the two possible shapes of elastic modulus curves. Both curves were obtained for samples containing Steol surfactant with W/C$_f$=0.41. Linear curve (empty circles) is obtained with 11.4~g/L of Steol and $\tau_{y,0}=$3~Pa, and non-linear curve (black dots), with 10.4~g/L of Steol and $\tau_{y,0}=$18~Pa.}
\label{graph_G_shape}
\end{center}
\end{figure}

Two different types of elastic modulus evolution are shown in figure \ref{graph_G_shape}. Curve slope can either be constant during 50~min or decrease with time.

\begin{figure}[!ht]
\begin{center}
\includegraphics[width=7cm]{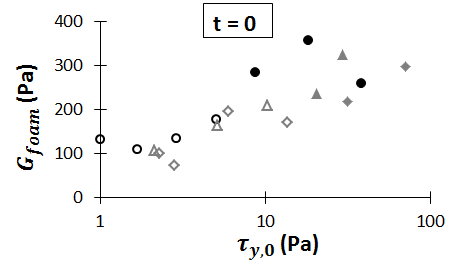}

\includegraphics[width=7cm]{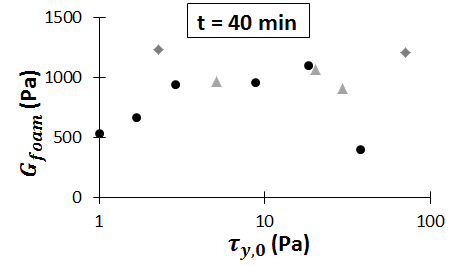}
\caption{Elastic modulus of fresh cement foam at t=0 and t=40~min. Black dots refer to cement pastes containing anionic surfactant at $W/C_f=0.41$ and grey diamonds at $W/C_f=0.38$. Grey triangles refer to samples with superplasticizer and non-ionic surfactant. In first graph, empty symbols refer to linear elastic modulus curves.}
\label{graph_Gfoam}
\end{center}
\end{figure}

Values for the elastic modulus at t=0 and t=40~min are plotted as a function of $\tau_{y,0}$ in Fig.~\ref{graph_Gfoam}. As expected, the higher $\tau_{y,0}$, the higher the initial elastic modulus. When reference cement paste yield stress increases by two decades, the foam elastic modulus is increased by a factor 4. However, 40 minutes after the start of the oscillation test, the foam elastic modulus hardly increases with reference cement paste yield stress anymore.

From Fig. \ref{graph_Gfoam}, we note that only samples with low $\tau_{y,0}$ correspond to samples characterized by a linear increase of the foam elastic modulus as a function time.

\subsection{Water suction}

\begin{figure}[!ht]
\begin{center}
\includegraphics[width=6cm]{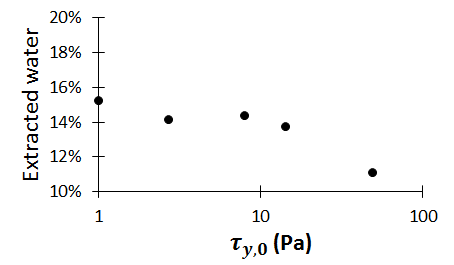}
\caption{Ratio of extracted water (i.e the volume of extracted water to the initial water volume) from reference cement pastes at $W/C_f=0.41$ containing Steol surfactant after 10 minutes at $\Delta P \simeq$ 400~Pa}
\label{Resultats_WaterExtraction}
\end{center}
\end{figure}

Fig.~\ref{Resultats_WaterExtraction} shows the percentage of water extracted from the reference cement pastes containing Steol at $W/C_f=0.41$. We note that the amount of extracted water decreases when $\tau_{y,0}$ increases, mainly when $\tau_{y,0}>\tau_{y,0}^*$. Below $\tau_{y,0}^*$, the volume ratio between the extracted water and the initial water is of the order of 15\%.

\section{Discussion}

\subsection{Comparison with simple aqueous foams}

Aqueous foams are known to behave like an elastic solid at small deformations and to exhibit a yield stress at higher deformation. Their rheological properties depend on the bubble radius $R$, the surface tension $\gamma$ and the gas volume content $\Phi$. Their yield stress is given by equation~\ref{equation_Tau_aq}~\cite{1996_Mason,1999_SaintJalmes} 
and their elastic modulus by equation~\ref{equation_Gaq}~\cite{2013_Cantat}.

\begin{equation}
\tau_{y,aq}(\Phi)=0.6 \dfrac{\gamma}{R}(\Phi-\Phi_c)^2
\label{equation_Tau_aq}
\end{equation}

\begin{equation}
G_{aq}(\Phi)=1.4 \dfrac{\gamma}{R}\Phi(\Phi-\Phi_c)
\label{equation_Gaq}
\end{equation}
where $\Phi_c$ is the critical packing volume fraction above which there are films between the bubbles and equals 0.64 in the case of disordered monodisperse foam.

To estimate both $\tau_{y,aq}$ and $G_{aq}$ the surface tension of bubbles $\gamma$ must be known. By considering the results published in~\cite{2017_Feneuil} for the surface tension of cement pore solutions, we use $\gamma = 35$ mN/m for both Steol or Triton foams. 

Therefore, using equations \ref{equation_Tau_aq} and \ref{equation_Gaq}, the aqueous foam yield stress for both surfactants is $\tau_{y,aq}(83\%)\simeq$ 2~Pa and the elastic modulus is $G_{aq}(83\%)\simeq$ 20~Pa. Both are significantly below the values measured with the cement foams samples for all $\tau_{y,0}$. This means that the interstitial cement paste strongly enhances the foam rheological properties, even when the reference cement paste yield stress $\tau_{y,0}$ is very low.

\subsection{Early age rheological properties}

Let us first discuss the presence of the overshoot when $\tau_{y,0} < \tau_{y,0}^*$. This overshoot is not expected to result directly from thixotropy effects \cite{2012_Roussel} in the cement paste, as they appear only for low reference cement paste yield stress, that is to say, for the highest concentrations of anionic surfactant or superplasticizer. Indeed, apparent thixotropy in cement paste arises from two effects: creation of a percolated network of colloidal cement grains during the first seconds after high shear mixing, then nucleation and growth of CSH bond between the grains~\cite{2012_Roussel}. Our yield stress measurements are carried out only a few minutes after the foam production, so the major contribution to the measured yield stress is the colloidal percolation network. As attraction forces between cement grains are reduced in the presence of high amount of Steol~\cite{2017_Feneuil} or superplasticizer~\cite{2012_Flatt} due to steric repulsion, the colloidal network is not expected to be stronger or to form faster in the case of deflocculated pastes.

We stress here that the start-of-flow curve of granular material exhibits an overshoot when grains are densely packed~\cite{1986_Bolton,2015_Fall}. Moreover, \textit{Gorlier et al.}~\cite{2017_Gorlier_4} studied start-of-flow curves of complex fluid foams. They showed that there is no overshoot when the interstitial fluid is a simple yield stress fluid (i.e. concentrated emulsion for example), whereas overshoot appears when aqueous foam is mixed with small beads suspension. The authors analyzed the stresses at stake in the granular foam and concluded that small particles packed in the foam structure behave as dense granular matter.

\begin{figure}[!ht]
\begin{center}
\includegraphics[width=7cm]{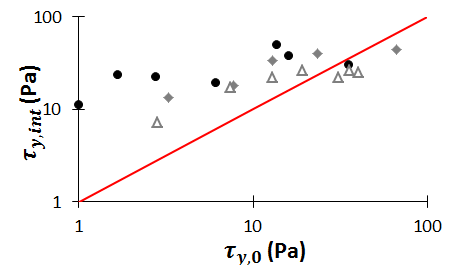}
\caption{Comparison of yield stress of identical cement pastes when they are confined in the foam $\tau_{y,int}$ and with no foam $\tau_{y,0}$. Black dots refer to cement pastes containing anionic surfactant at $W/C_f=0.41$ and grey diamonds at $W/C_f=0.38$. Grey triangles refer to samples with superplasticizer and non-ionic surfactant.}
\label{graph_interstitial_yieldstress}
\end{center}
\end{figure}

The yield stress of foams made with a yield stress fluid (characterized by a yield stress value $\tau_{y,int}$) has been shown to be described by equation~\ref{equation_Gorlier_YieldStress_c4}, provided that $\tau_{y,int}R/\gamma < 0.5$~\cite{2017_Gorlier_4}. 

\begin{equation}
\dfrac{\tau_{y,foam}(\Phi)}{\tau_{y,aq}(\Phi)}=1+c(1-\Phi)^{4/3}\left(\dfrac{\tau_{y,int}R}{\gamma} \right)^{2/3}
\label{equation_Gorlier_YieldStress_c4}
\end{equation}
where $c=110$ is a fitting parameter.
Using equation~\ref{equation_Gorlier_YieldStress_c4}, we deduce the yield stress of the interstitial cement paste $\tau_{y,int}$ from the measured foam mascroscopic yield stress $\tau_{y,foam}(\Phi)$ (plotted in Fig.~\ref{graph_foam_yieldstress}). $\tau_{y,int}$  is plotted in Fig.~\ref{graph_interstitial_yieldstress} against the reference cement paste yield stress $\tau_{y,0}$. This graph shows that, when the cement paste yield stress is high, i.e. above 20~Pa, the interstitial yield stress is equal to the reference cement paste yield stress. On the other hand, when the reference cement paste yield stress is lower, $\tau_{y,int}$ is strongly enhanced, up to a factor 10.

\bigbreak

To summarize, the start-of-flow curve shape (Fig.~\ref{graph_foam_start-of-flow}) and interstitial yield stress curves (Fig.~\ref{graph_interstitial_yieldstress}) allow us to identify two rheological behaviors of the cement paste in the foam:

\begin{itemize}
\item When $\tau_{y,0}>\tau_{y,0}^*$, the cement paste acts as a classical yield stress fluid, i.e. the measured value for the interstitial yield stress is close the yield stress of the reference cement paste. In this regime, that we call \textit{yield stress regime}, cement foams are unstable within our experimental conditions.
\item When $\tau_{y,0}<\tau_{y,0}^*$, the cement paste behaves as a confined granular material. An overshoot appears on the start-of-flow curve of the cement foam and the interstitial yield stress is significantly increased with respect to the reference cement paste yield stress. This regime, that we call \textit{granular regime}, because it is reminiscent of the behavior of confined granular packing, allows for significant foam stability. The stability loss at the lower reference cement paste yield stresses will be discussed in part~\ref{part_very_low_YS}.
\end{itemize}

\bigbreak

\begin{figure}[!ht]
\begin{center}
\includegraphics[width=7cm]{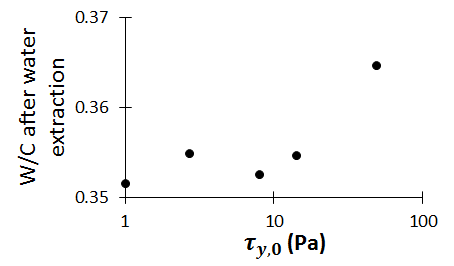}
\caption{Ratio of extracted water from reference cement pastes at $W/C_f=0.41$ containing Steol surfactant after 10 minutes at $\Delta P \simeq$ 400~Pa}
\label{WaterExtraction}
\end{center}
\end{figure}

For the \textit{granular regime} to appear, the cement grains packing must become denser, i.e. excess water must be removed. Estimation of the ability of cement paste to loose this excess water due to gravity effect is assessed by the water extraction curve. In Fig.~\ref{WaterExtraction}, we observe that, for a reference cement paste characterized by a water-to-cement ratio $W/C_f$=0.41, W/C reaches 0.35 when $\tau_{y,0}<\tau_{y,0}^*$. On the other hand, in the \textit{yield stress regime}, W/C remains above 0.36. This observation tends to confirm that, thanks to grains rearrangement, a denser granular network can be set up more easily when $\tau_{y,0}<\tau_{y,0}^*$ than when $\tau_{y,0}>\tau_{y,0}^*$. In cement foams, removal of water from the interstitial cement paste by drainage leads to an increase of its solid volume content $\Phi_p$. We recall that the yield stress of a solid suspensions is related to the interaction between the solid particles and the solid volume content \cite{2006_Flatt}. At given interaction intensity, the yield stress increases when the solid volume content increases. Therefore, water extraction due to drainage can account for the increase of $\tau_{y,int}$.

Let us estimate the amount of water which must drain from the  cement foams when water-to-cement ratio of the interstitial cement paste decreases from 0.41 to 0.35. Corresponding solid volume content in the cement paste is given by $\Phi_p=(1+\rho_c/\rho_{liq} W/C)^{-1}$; that is to say that $\Phi_p$ increases from 0.43 to 0.47. Water loss is therefore $(0.47-0.43)/0.47=8.5\%$ of the paste volume, i.e. 2\% of the foam volume, which cannot be visually observed with our samples.

\bigbreak

When $\tau_{y,0}<\tau_{y,0}^*$, granular effect can account for the increase of the yield stress of the interstitial cement paste due to the confinement provided by the bubble interface. However, interstitial yield stress is still smaller in the \textit{granular regime} than in the \textit{yield stress regime}, so other effects are involved in the remarkable stability observed in the \textit{granular regime}. One hypothesis to explain this stability is that the time evolution of the interstitial cement paste rheological properties plays a role on foam stability. This issue is discussed in the following paragraph.

\subsection{Time evolution of rheological properties}

Similarly to what we have done for $\tau_{y,int}$, we can estimate the interstitial elastic modulus $G_{int}$ from the measured value of the foam elastic modulus $G_{foam}$. The equation and the results are given and discussed in Appendix A. However, the interstitial elastic modulus, up to 40 min after foam preparation, cannot explain the remarkable stability of the foams in the \textit{granular regime}.

\begin{figure}[!ht]
\begin{center}
\includegraphics[width=7cm]{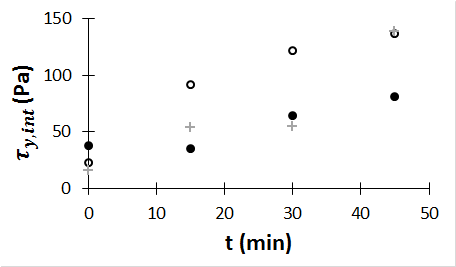}
\caption{Increase of interstitial yield stress with time for cement foams $W/C_i=0.35$ containing Steol. Full dots to foams in the colloidal regime (Steol concentration 10.4~g/L, $\tau_{y,0}=18~Pa$), empty dots refer to very stable foams in the granular regime (Steol concentration 11.4~g/L, $\tau_{y,0}=3~Pa$) and grey crosses to less stable foams in the granular regime (Steol concentration 12.3~g/L, $\tau_{y,0}< 1~Pa$)}
\label{graph_interstitial_yieldstress_time}
\end{center}
\end{figure}

Therefore, to further investigate the effect of foam aging, we measure the evolution of the interstitial yield stress with time. Whereas the elastic modulus is measured at low deformation, below 0.1\%, the yield stress is obtained at high shear strain. Consequently, elasticity and yield stress measured in cement paste have different origins~\cite{2012_Roussel}: the elasticity is caused only by the hydrates bonds between cement grains, whereas the yield stress results from both colloidal interaction and hydrate bonds. Second consequence of the high sample deformation during start-of-flow experiment is that this measurement is destructive: yield stress for each age must be measured with a different sample. Each sample is prepared following the same protocol as described in~\ref{section_protocol_c4} and placed in the rheometer geometry. Then a resting time between 0 and 45~min is chosen before the start of the yield stress measurement. Measured values for $\tau_{y,int}$ are shown in Fig.~\ref{graph_interstitial_yieldstress_time}. The curve shape in \textit{yield stress regime} is typical for cement paste~\cite{2012_Roussel} and exhibits a two-regime behavior with two different slopes: in the first regime, the yield stress growth is governed by the colloidal interactions between the cement grains, and, in the second regime, formation of hydrates bonds become predominant. 

For the stable foam in the \textit{granular regime}, a very fast increase occurs during the first minutes of rest. 15~minutes after the start of the experiment, $\tau_{y,int}$ is two times larger than the value obtained in the \textit{yield stress regime} and it appears to be able to prevent the foam destabilization. The curve shows that the drainage of the water from the interstitial cement paste, which is expected to induce the observed increase of $\tau_{y,int}$ within the granular regime, has not fully occurred when the initial rheological properties of the foams are measured (at t=0), but it seems completed after 15 min. After 15~minutes, both interstitial yield stress curves have the same slope, which shows that there is no major difference of chemical activity, and hence of hydrates nucleation, between the two types of samples.

When drainage is completed, i.e. after about 15~min, the value of $\tau_{y,int}$ obtained in the \textit{granular regime} is about 100~Pa. Note that this value is of the same order of magnitude as the interstitial yield stress obtained for foams made with  non-colloidal monodisperse spherical particles (diameter 10 or 20 $\mu$m)~\cite{2017_Gorlier_4}, i.e. 120~Pa. This confirms that the interstitial cement paste in the \textit{granular regime} behaves in a similar manner as a suspension of particles with very low attractive interactions.

\subsection{Stability loss at very high surfactant content}
\label{part_very_low_YS}

The last point that remains to be discussed is the stability loss observed for Steol foams when $\tau_{y,0}<\tau_{y,0}^{**}$. The evolution of $\tau_{y,int}$ for a foam where $\tau_{y,0} <$ 1~Pa is plotted in Fig.~\ref{graph_interstitial_yieldstress_time}. The increase of $\tau_{y,int}$ before 15~min is smaller than in the stable foam, but higher than in the case $\tau_{y,0}>\tau_{y,0}^{*}$.

\begin{figure}[!ht]
\begin{center}
\includegraphics[width=6cm]{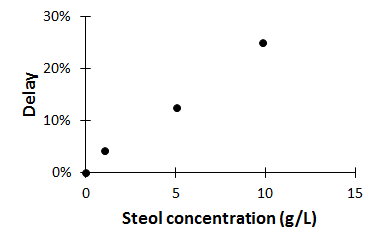}
\caption{Delay of hydration of cement pastes containing Steol. The paste are prepared following the same protocol as described in \cite{2017_Feneuil}, i.e. water-to-cement ratio is 0.5 and cement is a CEM I cement from St Vigor factory, Lafarge. We measure the evolution of temperature in the cement paste after water and cement mixing and the delay of hydration $\Delta$ is calculated from the time $t_{peak}$ when the maximum temperature is measured, with the formula:  $\Delta=(t_{peak}-t_{peak,ref})/t_{peak,ref}$. $t_{peak,ref}= 12~h$ is the reference time in sample containing no surfactant.}
\label{graph_delay_Steol}
\end{center}
\end{figure}

A first hypothesis to explain this stability loss is a delay of hydration. Indeed, some authors~\cite{2014_Wei} have observed that some surfactants, at high concentrations, delay hydration. This is confirmed for Steol in Fig.~\ref{graph_delay_Steol}. However, as mentioned in the previous paragraph, the influence of hydration kinetics on cement foam stability appears to be be minor compared to the early age structuration of the cement paste into a dense granular packing or colloidal network. 

Therefore, the most probable explanation for the observed unstability when $\tau_{y,int} < \tau_{y,int}^{**}$ is that drainage of the cement paste can occur between the bubbles. Indeed, a layer of cement can often be seen at the bottom of the highly deflocculated cement foam sample. The flow of a yield stress fluid in the foam channels, called Plateau borders, and nodes can be compared to the flow of a fluid in a porous medium. In such a case, the gravity-induced flow is expected to occur if the yield stress is below a critical value $\tau_{c,d}$ with depends on the pore radius $a$~\cite{1992_Chaplain}:

\begin{equation}
\tau_{c,d} \simeq  \rho g a
\label{equation_YS_drainage}
\end{equation}

In the case of a foam, we can assume that the equivalent pore size is represented by the size of the bigger spheres which can pass in the plateau borders. The corresponding radius is given by~\cite{2010_Louvet}:

\begin{equation}
r_{PB}= R \dfrac{0.27\sqrt{1-\Phi}+3.17(1-\Phi)^{2.75}}{1+0.57(1-\Phi)^{0.27}}
\label{equation_BdP}
\end{equation}

When R=390~$\mu$m and $\Phi$ = 83\%, equation~\ref{equation_BdP} gives $r_{PB}$= 40~$\mu$m. Therefore, equation \ref{equation_YS_drainage} gives $\tau_{c,d}\sim 1~Pa$, which is in agreement with our observation that the drainage occurs when $\tau_{y,0}<\tau_{y,0}^{**}=1 Pa$.

Note that drainage in the Plateau borders and nodes may induce a segregation of the cement grains in the very deflocculated cement pastes. The diameter of the biggest sphere that can pass in the Plateau borders is 2$r_{PB}$=80~$\mu$m, which is close to the size of the bigger cement grains. Therefore, while big cement grains are potentially retained by the foam nodes and channels, small grains can escape the structure and settle at the sample bottom. This leads to a less dense granular packing and a decreased interstitial yield stress.

\section{Conclusion}

We have studied the capacity of cement pastes for producing morphology-controlled cement foams. By using two distinct additives, our experimental approach allowed us to tune finely the reference cement paste yield stress (i.e. the yield stress of the unfoamed cement paste with the same water and additives content as the cement foam), while keeping constant several control parameters, namely the W/C ratio, the bubble size and the gas volume fraction. Our results reveal an unexpected effect of the reference cement paste yield stress on cement foam stability.

When the reference cement paste yield stress value is high, i.e. above 10~Pa, cement foams were found to evolve significantly before setting, leading to uncontrolled final morphology. In such a case we have shown that the interstitial cement paste behaves as a “classical” yield stress material with effective yield stress value equal to the yield stress of the reference cement paste. Note that within our experimental conditions, all studied reference cement paste yield stress values were below 100 Pa. We anticipate that larger values should allow for a control of the final morphology.

For lower reference cement paste yield stress values, remarkable morphological control can be achieved. This result is attributed to the additives-induced reduction of attractive van der Waals interactions: weak attractive forces allow for the densification of the cement grains within the foam network as well as the simultaneous drainage of the excess interstitial pore solution. The combination of these effects has been proved to enhance drastically the effective yield stress property of the interstitial cement paste, resulting in efficient immobilization of fresh foams, without resorting to set accelerators. Such a mechanical behavior is reminiscent of aqueous foams made with granular matter, i.e. grains without any other interaction than contacts, as studied recently by \textit{Gorlier et al.}~\cite{2017_Gorlier_4}. This result shows that, in practice, stable cement foams can be produced when cement paste yield stress is low. We anticipate that the drainage of the excess interstitial pore solution could be critical if the drained volume is large, i.e. for large W/C ratios or important foam heights.

Finally, we show that, when the reference cement paste yield stress is very low, poor control of the foam morphology is achieved. This result is attributed to the deflocculated state of cement colloidal particles in this regime: deflocculated small cement particles exit the foam skeleton along with the draining pore solution so the above-described densification mechanism is prevented from occurring. 

All these results are obtained with a constant bubble radius (390 $\mu$m).  A decrease of the bubble radius (i.e. a decrease of Plateau border size) is expected to better prevent the loss of cement grains from the interstitial cement paste. This would benefit to stability at very low reference cement paste yield stress. However, it would also increase ripening velocity \cite{2013_Cantat}.
Therefore, to understand the effect of bubble size on cement foam stability, the destabilization mechanisms at stake must be identified. This issue will be addressed in a further study. 

\section*{Acknowledgments}

The authors wish to thank David Hautemayou and Cédric Mézières for technical support. This work has benefited from two French government Grants managed by the Agence Nationale de la Recherche [Grants number ANR-11-LABX-022-01 and ANR-13-RMNP-0003-01].

\section*{Appendix A: Interstitial elastic modulus at 40 min.}

The foam elastic modulus $G_{foam}$ can be considered as the sum of the contribution of the aqueous foam $ G_{aq} $, and a contribution which depends on the interstitial elastic modulus $G_{int}$~\cite{1997_Gibson,2017_Gorlier_3}:

\begin{equation}
G_{foam}=G_{aq}+G_{int}(1-\Phi)^2\left(1+15 (2\Phi-1) \left(\dfrac{\gamma}{R G_{int}}\right)^{2/3}  \right)
\label{equation_Gorlier_G}
\end{equation}

\begin{figure}[!ht]
\begin{center}
\includegraphics[width=7cm]{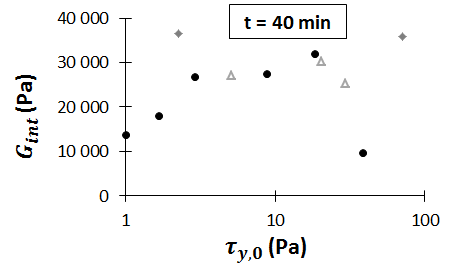}
\caption{Interstitial elastic modulus at time t=40~min. Black dots refer to cement pastes containing anionic surfactant at $W/C_f=0.41$ and grey diamonds at $W/C_f=0.38$. Grey triangles refer to samples with superplasticizer and non-ionic surfactant.}
\label{graph_interstitial_G}
\end{center}
\end{figure}

Equations~\ref{equation_Gaq} and \ref{equation_Gorlier_G} allow us to assess $G_{int}$ at several times, as shown in Fig.~\ref{graph_interstitial_G} for t=40~min.

We see that the increase of $G_{int}$ is faster in the \textit{granular regime} than in the \textit{yield stress regime}. Though, the elastic modulus remains higher in most of the \textit{yield stress}-type foams than in the \textit{granular regime} until 40~minutes. At this time, as shown in Fig.~\ref{pictures_destabilisation}, ripening has already started to occur in unstable foams. Therefore, the values of $G_{int}$ up to 40 minutes after foam production cannot account for the better stability observed in the \textit{granular regime}.

Regarding the shape of the curves of the elastic modulus as a function of time, we note that for foams in the \textit{granular regime}, the slope of foam elastic modulus curve is constant. This linear increase at low amplitude oscillations is a typical evolution for cement pastes and it is known to result from the constant volume formation rate of hydrates between the cement grains~\cite{2012_Roussel}. In the \textit{yield stress regime}, however, elastic modulus slope becomes smaller with time. We can assume that this is a consequence of the foam destabilisation: the slow shearing induced by the bubbles deformation during destabilization leads to a partial rejuvenation of the cement paste and partially compensates the increase of elasticity caused by thixotropy \cite{2012_Roussel}.

\bibliographystyle{elsarticle-num} 
\bibliography{Bibliographie}

\end{document}